\documentclass[11pt,a4paper]{article}
\usepackage{amsmath}
\usepackage{cite}
\usepackage{graphicx}
\usepackage{amsfonts}
\usepackage{amssymb}

\def\bt{
\beta}

\def\bt'{\beta'
}

\def\al{
\alpha }

\def\Ga{
\Gamma_{b} }

\def\Fus#1#2#3#4#5#6{
F_{#5#6}\left[
\begin{array}
[c]{cc}%
#3 & #2\\
#4 & #1%
\end{array}
\right]}

\begin{document}
\bigskip
\hfill\hbox{SISSA 82/2003/FM} \vspace{2cm}

\begin{center}
{\Large \textbf{Liouville Theory on the Pseudosphere:\\ \vspace{0.5cm}  Bulk-Boundary Structure Constant}} \\
\vspace{1.2cm} {\Large B\'en\'edicte Ponsot \footnote{\textsf
{ponsot@fm.sissa.it}}} \\
\vspace{0.7cm} {\it  International School for Advanced Studies (SISSA),\\
Via Beirut 2-4, 34014 Trieste, Italy}\\
\vspace{1.2cm}
\end{center}

\begin{abstract}
Liouville field theory on the pseudosphere is considered (Dirichlet conditions). We compute explicitly the bulk-boundary structure constant with two different methods: first we use a suggestion made by Hosomichi in JHEP 0111 (2001) that relates this quantity directly to the bulk-boundary structure constant with Neumann conditions, then we do a direct computation. Agreement is found.

\end{abstract}

\begin{center}
\end{center}

\vspace{1cm}
\section{Introduction}
We introduce the action density of the Liouville field theory
\begin{equation}
\mathcal{L}(z,\bar{z})= \frac{1}{4\pi}(\partial_{a}%
\phi(z,\bar{z}))^{2}+\mu e^{2b\phi(z,\bar{z})},
\nonumber
\end{equation}
where $\phi$ is the Liouville field, $\mu$ is called the cosmological constant and the parameter $b$ is the coupling constant.
LFT is a conformal field theory with central charge
$$
c_L=1+6Q^2\; ,
$$
where $Q=b+1/b$ is called the background charge. In what follows we will consider the so called weak coupling regime with $c_L \ge 25$.
We note the
conformal primary fields $V_{\al}(z,\bar{z})$. These fields
 are
primaries with respect to the energy momentum tensor
\begin{eqnarray}
T(z)&=&-(\partial\phi)^{2}+Q\partial^{2}\phi \; ,\nonumber \\
\bar{T}(\bar{z})&=&-(\bar{\partial}\phi)^{2}+Q\bar{\partial}^{2}\phi \; ,
\nonumber
\end{eqnarray}
and have conformal
weight $\Delta_{\alpha}=\bar{\Delta}_{\alpha}=\alpha(Q-\alpha)$.
One identifies the
operator $V_{\al}$ with its reflected image $V_{Q-\al}$:
\begin{eqnarray}
V_{\al}(z,\bar{z})=S(\al)V_{Q-\al}(z,\bar{z})
\label{reflection}
\end{eqnarray}
where we introduced the bulk reflection amplitude \cite{Dorn,ZZ}
$$
S(\al)=\frac{(\pi \mu
\gamma(b^{2}))^{(Q-2\al)/b}}{b^{2}}\frac{\gamma(2\al
b-b^{2})}{\gamma(2-2\al/b+1/b^{2})},
$$
which is unitary: $S(\al)S(Q-\al)=1$, and as usual 
$\gamma(x)=\Gamma(x)/\Gamma(1-x)$.\\
An important set among the
primaries are the fields $V_{-nb/2},\;n \in \mathbb{N},$ which are
degenerate with respect to the conformal symmetry algebra and
satisfy linear differential equations \cite{BPZ}. For example, the
first non trivial case consists of $\al=-b/2$, and the
corresponding operator satisfies
\begin{equation}
\left(\frac{1}{b^{2}}\partial^{2}+T(z)\right)V_{-b/2}=0, \nonumber
\end{equation}
as well as the complex conjugate equation.\\
It follows from these equations that when one performs the
operator product expansion of one of these degenerate operators
with a generic operator, then the OPE is truncated \cite{BPZ}. For
example:
\begin{eqnarray}
V_{-b/2}V_{\al}=c_{+}V_{\al-b/2}+c_{-}V_{\al+b/2} \, .\label{OPE}
\nonumber
\end{eqnarray}
The structure constants $c_{\pm}$ are special cases of the bulk
three point function, and can be computed perturbatively as
Coulomb gas (or screening) integrals \cite{FF,DF}. One can take
$c_+=1$, as in this case there is no need of insertion of
interaction, whereas $c_-$ requires one insertion of the Liouville
interaction $-\mu\int e^{2b\phi}d^2z $. It is given by the expression
$$
c_{-}=
-\mu\frac{\pi\gamma(2b\alpha-1-b^2)}{\gamma(-b^2)\gamma(2b\al)} \,
.\nonumber
$$
Similarly, there exists also a dual series of degenerate operators
$V_{-m/2b}$ with the same properties.\\
Liouville field theory on a pseudosphere was considered in \cite{ZZ2}\footnote{See also prior works on the subject \cite{DJ,DJ2}.}: the pseudosphere geometry can be realized as the disk $|z|<1$ with metric
$$
ds^2=e^{\phi(z)}|dz|^2,
$$
and 
$$
e^{\phi(z)}=\frac{4R^2}{(1-z\bar{z})^2},
$$
where $R$ is interpreted as the radius of the pseudosphere.
It was found in \cite{ZZ2} that there is a whole variety of vacua considered as boundary conditions at the absolute that are labelled by two positive integers $(m,n)$, in one to one correspondence with the degenerate representations of the Virasoro algebra. Then, the (finite) content of boundary operators
is simply determined by the fusion algebra, like in the rational case.
We will need later the expression for the one point function of a primary field $V_{\al}$ \cite{ZZ2}: 
\begin{eqnarray}
U_{m,n}(\al)=\frac{\sin(\pi b^{-1}Q)\sin (\pi m b^{-1}(2\al-Q))\sin(\pi bQ)\sin (\pi n b(2\al-Q))}{\sin(\pi mb^{-1}Q)\sin (\pi b^{-1}(2\al-Q))\sin(\pi nbQ)\sin (\pi  b(2\al-Q))}U_{1,1}(\al)
\label{onepoint}
\end{eqnarray}
where the $(1,1)$ one point function reads
\begin{eqnarray}
U_{1,1}(\al)=\frac{(\pi \mu \gamma(b^2))^{-\al/b}\Gamma(bQ)\Gamma(Q/b)Q}{\Gamma(bQ-2b\al)\Gamma(Q/b-2\al /b)(Q-2\al)}
\end{eqnarray}
This expression satisfies the reflection property (\ref{reflection}).

\section{Hosomichi's proposal \cite{hosomichi}}
It was first noticed in \cite{ZZ2} that the one point function in the pseudosphere geometry (\ref{onepoint}) is related to the one point function with Neumann boundary conditions computed in \cite{FZZ}. The latter result has the following expression:
\begin{eqnarray}
U_{s}(\alpha)=\frac{2}{b}(\pi \mu
\gamma(b^2))^{\frac{(Q-2\alpha)}{2b}} \Gamma(2b\alpha-b^2)\Gamma(2b^{-1}\alpha-b^{-2}-1)
\cosh[2\pi(2\alpha-Q)s],\nonumber
\label{onepointN}
\end{eqnarray}
where the boundary parameter $s$ is related to the boundary cosmological constant \cite{FZZ}:
\begin{eqnarray}
\text{cosh}\left(2\pi
bs\right)=\frac{\mu_{B}}{\sqrt{\mu}}\sqrt{\text{sin}(\pi
b^{2})}. \nonumber 
\label{relation mu-sigma}
\end{eqnarray}
Let us note that (\ref{onepointN}) satisfies the reflection
 property (\ref{reflection}).\\
If one first perform a Fourier transform w.r.t the parameter $s$ on the one point function:
\begin{eqnarray}
\tilde{U}(\alpha,p)=\frac{1}{2}\int_{-\infty}^{+\infty}e^{4\pi sp}U_{s}(\alpha)\;ds
\label{fourier}
\end{eqnarray} 
and then one does the transformation:
\begin{eqnarray}
\int_{-i\infty}^{+i\infty}\sin(2\pi npb)\sin(2\pi mpb^{-1})\tilde{U}(\alpha,p)\; dp,
\label{transfo}
\end{eqnarray}
this reproduces, up to the factor
\begin{eqnarray}
-(\pi\mu\gamma(b^{2}))^{-Q/2b}\frac{\sin(\pi b^{-1}Q)\sin(\pi bQ)}{\sin(\pi mb^{-1}Q)\sin(\pi nbQ)}\Gamma(bQ)\Gamma(Q/b)Q
\label{factor}
\end{eqnarray}
the expression (\ref{onepoint}). It was then proposed in \cite{hosomichi} that this relation also holds at the level of the bulk-boundary structure constant, which is what we are going to check.\\
The bulk-boundary structure constant with Neumann boundary conditions was calculated in \cite{hosomichi}; it has the form of a $b$-deformed hypergeometric function in the Barnes representation:
\begin{eqnarray}
\lefteqn{R_{s}(\alpha,\beta)=2\pi(\pi\mu\gamma(b^{2})b^{2-2b^{2}})^{\frac{1}{2b}(Q-2\alpha-\beta)}}\nonumber
\\
&&\times \quad \frac{\Gamma_b^3(Q-\beta)\Gamma_b(2\alpha-\beta)\Gamma_b(2Q-2\alpha-\beta)}{\Gamma_b(Q)\Gamma_b(Q-2\beta)\Gamma_b(\beta)
\Gamma_b(2\alpha)\Gamma_b(Q-2\alpha)}\nonumber \\
&& \times \quad \int_{-i\infty}^{i\infty} dt\; e^{-4\pi
ts}\frac{S_b(t+\beta/2+\alpha-Q/2)S_b(t+\beta/2-\alpha+Q/2)}
{S_b(t-\beta/2-\alpha+3Q/2)S_b(t-\beta/2+\alpha+Q/2)}.\nonumber \\
\label{bb}
\end{eqnarray}
We introduced the Barnes' Double Gamma function:
\begin{eqnarray}
\text{log}\Gamma_{2}(s|\omega_1,\omega_2)=\left(\frac{\partial}{\partial
t}
\sum_{n_1,n_2=0}^{\infty}(s+n_1\omega_1+n_2\omega_2)^{-t}\right)_{t=0},
\nonumber
\end{eqnarray}
and by definition $\Gamma_b(x) \equiv \frac{\Gamma_2(x|b,b^{-1})}{\Gamma_2(Q/2|b,b^{-1})}$. This function satisfies the functional relation
$\Ga(x+b)= \frac{\sqrt{2\pi}b^{bx-\frac{1}{2}}}{\Gamma(bx)}\Ga(x)$, as well as the dual relation with $b$ replaced by $b^{-1}$.
$\Ga(x)$ is a meromorphic function of $x$, whose poles are located
at
$x=-nb-mb^{-1}, n,m \in \mathbb{N}$. The $S_b(x)$ function is related to the $\Gamma_b(x)$ function: $S_b(x)\equiv \frac{\Ga(x)}{\Ga(Q-x)}$.\\
The integration contour of the integral (\ref{bb}) is located to the right of the poles:
\begin{eqnarray}
t = -\beta/2-\alpha+Q/2-\nu b-\mu b^{-1},\quad 
t = -\beta/2+\alpha-Q/2-\nu b-\mu b^{-1},\quad  \mu,\nu \in \mathbb{N}, 
\nonumber
\end{eqnarray}
and to the left of the poles:
\begin{eqnarray}
t = \beta/2+\alpha-Q/2 +\nu b+\mu b^{-1},\quad
t = \beta/2-\alpha+Q/2+\nu b+\mu b^{-1},\quad \mu,\nu \in \mathbb{N}.
\nonumber
\end{eqnarray}
Let us consider now a degenerate boundary operator $B_{\beta}^{\sigma\sigma}$ with spin $\beta=-ub-vb^{-1}$, $u$ and $v$ being  positive integers. What should be seen is that for this value of $\beta$, we have to pick up residues at poles\footnote{This computation does not hold for the case $2\al=\beta$; as it was noticed in \cite{KPS}, the $b$-deformed hypergeometric function degenerates for this value.} that are located at
\begin{eqnarray}
t=\pm(\alpha+\beta/2-Q/2+kb+lb^{-1}),\quad k=0,\dots,u, \quad l=0,\dots,v.
\label{residues}
\end{eqnarray}
It is convenient to introduce at this point the truncated $b$-deformed hypergeometric series:
\begin{eqnarray}
\phi (A=-ub,B;C;-ix) = \sum_{k=0}^{u}e^{2\pi kbx}\prod_{i=0}^{k-1}
\frac{\sin \pi b(B+ib)\sin \pi b(A+ib)}{\sin \pi b(C+ib)\sin \pi b (Q+ib)},
\nonumber
\end{eqnarray}
and define
\begin{eqnarray}
\phi_b^u (x)\equiv \phi(-ub,2\alpha-ub-Q;2\alpha;-ix).
\nonumber
\end{eqnarray}
The evaluation of the residues (\ref{residues}) gives
\begin{eqnarray}
\lefteqn{\frac{1}{b}(\pi\mu\gamma(b^{2}))^{\frac{1}{2b}(Q-2\alpha-\beta)}\;
\Gamma(2b\alpha-b^2)\Gamma(2b^{-1}\alpha-1-b^{-2})  b^{b^2-b^{-2}}\quad \times }\nonumber \\
&& \prod_{i=0}^{u-1}\frac{\Gamma(2b\al-bQ+b\beta+ib^2)}{\Gamma(2b\al+ib^2)}\frac{\Gamma(bQ-b\beta+ib^2)}{\Gamma(bQ+ib^2)}\prod_{j=0}^{v-1}\frac{\Gamma(\frac{2\alpha}{b}-\frac{Q}{b}+\frac{\beta}{b}+\frac{j}{b^{2}}+u)}{\Gamma(\frac{2\alpha}{b}+\frac{j}{b^{2}}+u)}\frac{\Gamma(\frac{Q}{b}-\frac{\beta}{b}+\frac{j}{b^{2}}+u)}{\Gamma(\frac{Q}{b}+\frac{j}{b^{2}}+u)}
\nonumber \\
&&
\times \quad \left(e^{2\pi s(2\alpha+\beta-Q)}\phi_b^u (2s)\phi_{1/b}^v (2s)+\quad s\to -s \right).
\end{eqnarray}
Then, after performing transformations (\ref{fourier}), (\ref{transfo}) and multiplying by (\ref{factor}), we obtain, for $2\alpha \neq \beta$:
\begin{eqnarray}
\lefteqn{R_{m,n}(\alpha,\beta)=-\frac{ (\pi\mu\gamma(b^{2}))^{-\frac{\beta}{2b}}\; U_{m,n}(\alpha)}{4\sin \pi nb(2\al-Q)\sin \pi mb^{-1}(2\al-Q)}\; b^{b^2-b^{-2}}
\times }\nonumber \\
&& \prod_{i=0}^{u-1}\frac{\Gamma(2b\al-bQ+b\beta+ib^2)}{\Gamma(2b\al+ib^2)}\frac{\Gamma(bQ-b\beta+ib^2)}{\Gamma(bQ+ib^2)}\prod_{j=0}^{v-1}\frac{\Gamma(\frac{2\alpha}{b}-\frac{Q}{b}+\frac{\beta}{b}+\frac{j}{b^{2}}+u)}{\Gamma(\frac{2\alpha}{b}+\frac{j}{b^{2}}+u)}\frac{\Gamma(\frac{Q}{b}-\frac{\beta}{b}+\frac{j}{b^{2}}+u)}{\Gamma(\frac{Q}{b}+\frac{j}{b^{2}}+u)}
\nonumber \\
&& \times \quad \left( e^{\sqrt{-1}\pi nb(2\alpha+\beta-Q)}\phi_b^u(\sqrt{-1}nb) - \quad n\to -n\right)
\left(  e^{\sqrt{-1}\pi mb^{-1}(2\alpha+\beta-Q)}\phi_{1/b}^v(\sqrt{-1}mb^{-1}) - \quad m\to -m\right)\; .\nonumber \\
\label{bbpseudo}
\end{eqnarray}
One can check that this expression satisfies the reflection property (\ref{reflection}).\\
It is not difficult to see that for $n=1$ and $\beta=-b$ as well as for $n=1,2$ and $\beta=-2b$, the term in the first parenthesis vanishes identically. We have no doubt, although we did not do it in general, that our coefficient $R_{m,n}(\alpha,-ub-vb^{-1})$ is zero\footnote{Vanishing of the structure constant will always be due to the terms into parenthesis.} whenever the fusion rules for the degenerate representations corresponding to the boundary conditions and the boundary operator are not satisfied, as expected from the results of \cite{ZZ2}.

\section{Direct computation}
We use the trick of \cite{Teschner} and consider an auxiliary bulk two point function including a degenerate operator $V_{\beta/2}$ and a generic operator $V_{\alpha}$. For the sake of simplicity, we shall consider $\beta=-ub$.
The two point function can be factorized equivalently in the $s$- and $t$-channels. A straightforward generalization of the case $u=1$ already studied in \cite{ZZ2} leads to the following equation for $R_{m,n}(\al,\beta)$\footnote{Such an equation was first obtained in \cite{CL} for the case of A-type Virasoro minimal models.}\footnote{When the bulk operator $V_{-ub/2}$ approaches the boundary, it gives rise to primary boundary operators $B_{0},B_{-b},\cdots,B_{-ub}$. We consider here the contribution of $B_{-ub}\equiv B_{\beta}$.}:
\begin{align}
\sum_{k=0}^{u}C(\alpha,\beta/2,Q-\al-\beta/2-kb)U_{m,n}(\al+\beta/2+kb)\Fus{\al}{\beta/2}{\beta/2}{\al}{\al+\beta/2+kb,}{\beta}  \nonumber \\
= R_{m,n}(\al,\beta)R_{m,n}(\beta/2,\beta)D_{m,n}(\beta)\nonumber
\end{align}
where we introduced:
\begin{itemize}
\item
The bulk three point function $C(\al_1,\al_2,\al_3)$. In the case where $\al_1+\al_2+\al_3=Q-kb$, its value can be found in \cite{DF}:
\begin{eqnarray}
C(\al_1,\al_2,\al_3)=\left(\frac{-\pi \mu}{\gamma(-b^2)}\right)^k
\frac{\prod_{i=1}^{k}\gamma(-ib^2)}{\prod_{i=0}^{k-1}[\gamma(2b\al_1+ib^2)
 \gamma(2b\al_2+ib^2)\gamma(2b\al_3+ib^2)]}.\nonumber
\end{eqnarray}
\item
$U_{m,n}(\al)$ is defined as in equation (\ref{onepoint}).
\item
$D_{m,n}(\beta)$ is the boundary two point function of two degenerate boundary operators with spin $\beta$. It is usual in conformal field theory to normalize the two point function of primaries to one.
\item
$\Fus{\al}{\beta/2}{\beta/2}{\al}{\al+\beta/2+kb,}{\beta}$ is a special case of the fusion matrix, which expresses the change of basis between the $s$-channel conformal block and the $t$-channel conformal block. The fusion matrix was built for generic spins in \cite{PT1}; we recall its expression:
\begin{eqnarray}
\lefteqn{\Fus{\al_1}{\al_2}{\al_3}{\al_4}{\al_{21}}{\al_{32}}=} \nonumber \\
&&
\frac{\Ga(2Q-\al_3-\al_2-\al_{32})\Ga(\al_3+\al_{32}-\al_2)\Ga(Q-\al_2-\al_{32}+\al_3)\Ga(Q-\al_3-\al_2+\al_{32})}{\Ga(2Q-\al_1-\al_2-\al_{21})\Ga(\al_1+\al_{21}-\al_2)\Ga(Q-\al_2-\al_{21}+\al_1)\Ga(Q-\al_2-\al_1+\al_{21})} \nonumber \\
&&
\frac{\Ga(Q-\al_{32}-\al_1+\al_4)\Ga(\al_{32}+\al_{1}+\al_{4}-Q)\Ga(\al_1+\al_4-\al_{32})\Ga(\al_4+\al_{32}-\al_1)}{\Ga(Q-\al_{21}-\al_{3}+\al_4)\Ga(\al_{21}+\al_3+\al_4-Q)\Ga(\al_3+\al_4-\al_{21})\Ga(\al_{21}+\al_4-\al_3)} \nonumber \\
&&\frac{\Ga(2Q-2\al_{21})\Ga(2\al_{21})}{\Ga(Q-2\al_{32})\Ga(2\al_{32}-Q)}
\frac{1}{i}\int\limits_{-i\infty}^{i\infty}dt \;\;
\frac{S_b(U_1+t)S_b(U_2+t)S_b(U_3+t)S_b(U_4+t)}
{S_b(V_1+t)S_b(V_2+t)S_b(V_3+t)S_b(Q+t)} \nonumber
\label{formule
de F}
\end{eqnarray}
where:
$$
\begin{array}{ll}
 U_1 = \al_{21}+\al_1-\al_2                &  V_1 = Q+\al_{21}-\al_{32}-\al_{2}+\al_{4} \\
 U_2 = Q+\al_{21}-\al_2-\al_1              &  V_2 = \al_{21}+\al_{32}+\al_{4}-\al_2 \\
 U_3 = \al_{21}+\al_{3}+\al_{4}-Q          &  V_3 = 2\al_{21} \\
 U_4 = \al_{21}-\al_{3}+\al_4 \\              \\
\end{array}
$$
In the case of degenerate Virasoro representations, the generic fusion coefficient developes poles; the relevant quantity is given by the residue at these poles. In our case the pole is located at $t=Q-2\alpha-\beta-kb$. We find:
\begin{eqnarray}
\lefteqn{\Fus{\al}{\beta/2}{\beta/2}{\al}{\al+\beta/2+kb,}{\beta}=}\nonumber \\
&&\prod_{l=1}^{u}\Gamma(bQ-b\beta+(l-1)b^2)\prod_{l=k}^{u-1}\frac{\Gamma(2b\alpha+b\beta+(l+k)b^2)}{\Gamma(2b\alpha+lb^2)\Gamma(bQ+lb^2)}\nonumber \\
&&\times
\prod_{i=1}^{k}\frac{\Gamma(bQ+(i-1)b^2)\Gamma(2bQ-2b\al-b\beta-2kb^2+(i-1)b^2)}{\Gamma(bQ-b\beta-ib^2)^2\Gamma(2bQ-2b\al-b\beta-ib^2)}.\nonumber 
\end{eqnarray}
\end{itemize}
We checked that the expression found for $R_{m,n}(\al,\beta=-ub)$ with this method
indeed coincides with (\ref{bbpseudo}), provided $R_{m,n}(\beta/2,\beta)$ is normalized to one for those values of $m,n$ which make the bulk-boundary coefficient $R_{m,n}(\al,\beta)$ non vanishing.

\section*{Acknowledgments}
Work supported by the Euclid Network HPRN-CT-2002-00325.

\end{document}